# α7 nicotinic acetylcholine receptor signaling modulates ovine fetal brain astrocytes transcriptome in response to endotoxin

M. Cao[1*], J.W. MacDonald[2*], H.L. Liu[1], M. Weaver[3], M. Cortes[4], L.D. Durosier[1], P. Burns[5], G. Fecteau[5], A. Desrochers[5], J. Schulkin[6], M. C. Antonelli[7], Bernier, R.A.[8], M. Dorschner[3], T. K. Bammler[2], M.G. Frasch[1,4,6,9]

[1] Dept. of Obstetrics and Gynaecology and Dept. of Neurosciences, CHU Ste-Justine Research Centre, Faculty of Medicine, Montréal, QC, Canada;
[2] Dept. of Environmental and Occupational Health Sciences, University of Washington, Seattle, WA, USA
[3] UW Medicine Center for Precision Diagnostics, University of Washington, Seattle, WA, USA
[4] Animal Reproduction Research Centre (CRRA), Faculty of Veterinary Medicine, Université de Montréal, Montréal, QC, Canada;
[5] Dept. of Clinical Sciences, Faculty of Veterinary Medicine, Université de Montréal, QC, Canada;
[6] Dept. of Obstetrics and Gynecology, University of Washington, Seattle, WA, USA.
[7] Instituto de Biología Celular y Neurociencia "Prof. Eduardo De Robertis", Facultad de Medicina, Universidad de Buenos Aires, Argentina
[8] Department of Psychiatry and Behavioral Sciences, University of Washington, Seattle, WA
[9] Center on Human Development and Disability, University of Washington, Seattle, WA

* MC and JWM contributed equally to this paper.

**Running head**: Fetal astrocytes α7nAChR signaling

Address of correspondence:
Martin G. Frasch
Department of Obstetrics and Gynecology
University of Washington
1959 NE Pacific St
Box 356460
Seattle, WA 98195
Phone: +1-206-543-5892
Fax: +1-206-543-3915
Email: mfrasch@uw.edu

**ABSTRACT**

Neuroinflammation *in utero* may result in lifelong neurological disabilities. Astrocytes play a pivotal role, but the mechanisms are poorly understood. No early postnatal treatment strategies exist to enhance neuroprotective potential of astrocytes. We hypothesized that agonism on α7 nicotinic acetylcholine receptor (α7nAChR) in fetal astrocytes will augment their neuroprotective transcriptome profile, while the antagonistic stimulation of α7nAChR will achieve the opposite. Using an *in vivo - in vitro* model of developmental programming of neuroinflammation induced by lipopolysaccharide (LPS), we validated this hypothesis in primary fetal sheep astrocytes cultures re-exposed to LPS in the presence of a selective α7nAChR agonist or antagonist. Our RNAseq findings show that a pro-inflammatory astrocyte transcriptome phenotype acquired *in vitro* by LPS stimulation is reversed with α7nAChR agonistic stimulation. Conversely, antagonistic α7nAChR stimulation potentiates the pro-inflammatory astrocytic transcriptome phenotype. Furthermore, we conduct a secondary transcriptome analysis against the identical α7nAChR experiments in fetal sheep primary microglia cultures and discuss the implications for fetal and postnatal brain development.

**Introduction**

Glial cells (astrocytes and microglia) play a role in neuroinflammation and both cell types acquire a specific reactive phenotype when stimulated by lipopolysaccharide (LPS).(1) Activation of glial cells may lead to neuronal cell death. Activation of nicotinic α7 receptors (α7nAChR) suppresses the LPS-induced reactive phenotype of microglia and astrocytes and thus counteracts the deleterious effect regarding neuronal viability.(2–8)

In the periphery the efferent fibers of the Vagus nerve activate α7nAChR on effector cells and inhibit inflammation.(9) In the brain, fibers arising from the Nucleus tractus solitarii spread into both hemispheres and their activation may lead to a widespread central anti-inflammatory effect.(10,11)(5) It is currently insufficiently tested if this is also true for the fetal brain.

In a previous experiment we investigated the effect of α7nAChR stimulation on LPS-induced microglia-activation in a double hit model of sheep fetal microglia. In the current experiment, we extended this investigation to study the role of α7nAChR in fetal sheep astrocytes. These experiments may help to shed light on neurodevelopmental disorders such as autism spectrum disorder (ASD) or schizophrenia, that are thought to involve neuroinflammation during the fetal period.(12,13)

We hypothesized that (1) under exposure to LPS, α7nAChR agonist stimulation in fetal astrocytes augments their neuroprotective profile, while the antagonistic stimulation reduces it; (2) a LPS double-hit (first *in vivo*, then *in vitro*) on astrocytes exacerbates these effects similar to microglia as demonstrated before. Using an *in vivo - in vitro* fetal sheep model (14), we validate these hypotheses via RNASeq analysis in primary fetal astrocyte cultures exposed to LPS in the presence of a selective α7nAChR agonist or antagonist. We compare these findings to the previously published results in identically conducted microglia experiments.(3,15)

**Methods**

*Study approval*

This study was carried out in strict accordance with the recommendations in the Guide for the Care and Use of Laboratory Animals of the National Institutes of Health. The respective *in vivo* and *in vitro* protocols were approved by the Committee on the Ethics of Animal Experiments of the Université de Montréal (Permit Number: 10-Rech-1560).

*Astrocytes isolation and purification*

The detailed protocol has been presented elsewhere.(3) Briefly, fetal sheep brain tissues were obtained during sheep necropsy after completion of the *in vivo* experiment to conduct the *in vitro* study (Fig. 1). In the *in vivo* experiments, three *in utero* instrumented fetal sheep were treated intravenously with LPS (400ng/fetus/day) derived from E. coli (Sigma Cat. no L5293, E. coli O111:B4, ready-made stock solution at a concentration of 1mg/ml) on experimental days 1 and 2 at 10:00am to mimic high levels of endotoxin in fetal circulation (so-called first LPS exposure or first hit). Three *in utero* instrumented fetal sheep were used as control receiving sterile saline. The instrumented fetuses were referred to as primary fetuses. In case of twins, twin fetuses were not instrumented and their brains directly used for subsequent cell culture. Fetuses not exposed to LPS, either primary or twins, were designated "naïve" (no LPS exposure *in vivo*). Instrumented animals that received LPS *in vivo* were used for second hit LPS exposure *in vitro*.

Astrocytes are the major adherent cell population in flask. Astrocytes were purified by passage into a new T75 flask for 4-5 times before any manipulations and treatments. After floating microglia collection, the adherent cells were detached by trypsinization (Trypsine 0.25% + EDTA 0.1%, Wisent Cat. No 325-043-EL) and re-plated into a new flask. Cells were cultured for another 7 days with 10% ready-to-use medium (DMEM plus 1% penicillin/ streptomycin, 1% glutamine, in addition with 10% heat-inactivated fetal bovine serum (Gibco, Canada Origin)). The cell passage procedure took 4-5 weeks until purified astrocytes could be used for the *in vitro* experiment. The cell culture conditions were 37°C, 5% $CO_2$.

Pure astrocytes were plated into a 24-well plate at $1 \times 10^5$ cells/mL with 10% DMEM for another 7 days, and then treated with LPS or saline for 6h.

Cell-conditioned media were collected for cytokine analysis. To verify astrocytes purity, a portion of cells was plated into Lab-Tek 8 well chamber glass slide (Thermo Scientific) for immunocytochemistry (ICC) analysis. Glial fibrillary acidic protein (GFAP) was used as an astrocyte marker; cells were counterstained with Hoechst.(15)

*Astrocyte cell culture and treatment*

Prior to exposure to LPS (Sigma Cat. no L5024, E. coli O127:B8) at a concentration of 100 ng/ul, cells were pretreated for 1 hour with either 10nM AR-R17779 hydrochloride (Tocris Bioscience Cat# 3964), a selective α7nAChR agonist, or 100 nM α-Bungarotoxin (Tocris Bioscience Cat# 2133), a selective α7nAChR antagonist. Optimal dose of AR-R17779 (A) or α-Bungarotoxin (B) was chosen based on a dose-response experiment with LPS. We have tested 0, 10, 100, and 1000nM of α-Bungarotoxin and 0, 1, 10, and 100nM of AR-R17779 in the absence or presence of 100ng/ul LPS, and measured IL-1β concentrations in cultured media as the endpoint. The 100nM α-Bungarotoxin and 10nM AR-R17779 were chosen because the cells responded in a linear range as indicated by IL-1β production.

AR-R17779 was reconstituted in DMSO as stock solution, serial dilutions were made to prepare the working stock; to obtain 10nM AR-R17779 in concentration per well, 5ul working stock was added well by well containing 500ul media; only DMSO was added in control well, therefore, the DMSO concentration per well was 1%. α-Bungarotoxin was reconstituted with culture media into a stock solution and underwent serial dilutions.

In a complete cell culture experiment, we had four experimental groups: Control (naïve control), LPS100 (naïve LPS), LPS100+B10 (naïve LPS+B) and LPS100+A10 (naïve LPS+A). Second hit cell cultures were designed with the same pattern and divided into four experimental groups: Control (SHC), LPS100, LPS100+B100 and LPS100+A10.

*Measurement of the cytokine IL-1β in cell culture media*

The approach is described elsewhere.(14,16) Briefly, IL-1β concentrations in cell culture media was determined by using an ovine-specific sandwich ELISA. 96-well plates were pre-coated with the mouse anti sheep monoclonal antibodies (IL-1β, MCA1658, Bio Rad AbD Serotec) at a concentration of 4μg/ml on an ELISA plate at 4°C overnight. After 3 times wash with washing buffer (0.05% Tween 20 in PBS, PBST), plates were blocked for 1h with 1% BSA in PBST for plasma samples or 10% FBS for cell culture media. Recombinant sheep proteins (IL-1β, Protein Express Cat. no 968-405) were used as ELISA standard. All standards and samples were run in duplicate (50μl per well). Rabbit anti-sheep polyclonal antibodies (IL-1β, AHP423, Bio Rad AbD Serotec) at a concentration of 4μg/ml were applied in wells and incubated for 30 min at room temperature. Plates were washed with washing buffer for 5-7 times between each step. Detection was accomplished by assessing the conjugated enzyme activity (goat anti-rabbit IgG-HRP, dilution 1:5000, Jackson ImmunoResearch, Cat. No 111-035-144) via incubation with TMB substrate solution (BD OptEIA TMB substrate Reagent Set, BD Biosciences Cat. No 555214); colour development reaction was stopped with 2N sulphuric acid. Plates were read on an ELISA plate reader at 450 nm, with 570 nm wavelength correction (EnVision 2104 Multilabel Reader, Perkin Elmer). The sensitivity of IL-1β ELISA was 41.3 pg/ml. The intra-assay and inter-assay coefficients of variance were <5% and <10%, respectively.

*RNAseq approach*

Nine replicates of astrocyte culture were studied, with one control and three treatment groups (Ctrl, LPS100, LPS+A10, LPS+B100) per replicate. Four replicates were used for RNAseq based on the RNA quality, of which 3 replicates of naive, one replicate of second-hit, with one treatment missing in this second-hit replicate (agonist treatment), fifteen samples in total were assessed with RNAseq in this study (Table 1).

<u>RNA extraction and quantification</u>: We used Qiagen RNeasy Mini Kit (Cat no 74104) for RNA extraction. RNA quantity and quality (RNA integrity number, RIN) was established by using a RNA Nano Chip (Agilent RNA 6000 Nano Chips) with Agilent 2100 BioAnalyzer. All samples had an acceptable RIN value ranging from 8.4 to 9.6. A total of twelve naïve astrocyte cultures from four sets of replicates was selected for RNA sequencing at high throughput, as well as three second hit astrocyte cultures(Table 1).

RNA-seq libraries were prepared using the TruSeq stranded mRNA kit (Illumina cat #20020594) and quality control was performed on the Agilent TapeStation and using the KAPA SYBRFAST qPCR kit. Single read 100-bp sequencing was performed in rapid mode on a HiSeq 2500 at the University of Washington Northwest Clinical Genomics Lab (Department of Pathology).

*RNAseq data analyses*

*Astrocytes*

The goal for this analysis was to compare the single hit LPS treated samples to Control, as well as making comparisons between the LPS treated samples (e.g., testing for changes due to the additional agonist or antagonist pre-treatment). The second hit samples had no replicates, so we could only make the directed comparisons between the single and second hit samples. For example, comparing the single and second hit LPS treated samples, as before.(3)

We aligned the reads to the Oar_v3.1 transcriptome using the salmon aligner (17), which infers the most likely transcript for each read using a quasi-mapping algorithm. We then 'collapsed' the transcript read counts to the gene level by summing up the reads for each gene's transcripts, using the Bioconductor tximport package.(18) In the end, we had a set of read counts per gene, for each sample.

Four replicates were used for RNAseq, with one control and three treatment groups Ctrl, LPS, LPS+A10, LPS+B100) per replicate. There were 3 replicates of naïve, one replicate of second-hit, with one treatment missing in this second-hit replicate (second-hit + agonist, SHA), fifteen samples in total were assessed with RNAseq in this study (Table 1). The instrumented fetuses were designated as primary fetus, identified as animal ID+P (stands for primary), whereas animal ID+T stands for non-instrumented twins.

To compare astrocytes transcriptomes, we used the Bioconductor edgeR package (19), to fit a generalized linear model with a negative binomial link function, and made comparisons between groups using quasi-likelihood F-tests.

We fitted the model described above, including the treatment effect for each animal, and made the comparisons, incorporating a fold change criterion into the test. In other words, the conventional test for a difference is between the null hypothesis $H_0 : \beta = 0$ versus the alternative hypothesis $H_A : \beta \neq 0$, but this may include very small changes that are likely to be biologically meaningless. One alternative to exclude such genes is to use a *post hoc* fold change adjustment, where we select genes based on the observed fold change between groups. This is problematic because we ignore the imprecision in our estimate of fold change. A better method is to incorporate the fold change into our inference, where we test $H_0 : \beta < |c|$ against the alternative $H_A : \beta \geq |c|$, for some constant fold change. By doing this, we are testing to see if we have evidence that the underlying population differences are larger than a given fold change, rather than simply testing that our sample data fulfill those criteria. We used a 1.5-fold change, and a false discovery rate (FDR) of 0.05 as criteria to define significantly differentially expressed genes, meaning that we expect that there are, at most, 5% false positives in our set of significant genes.

*Microglia - astrocytes transcriptome comparisons*

We also made comparisons to the existing RNAseq data that our lab generated using primary cultures of fetal sheep microglia from the identical experimental design in three biological replicates. This data has been published and the data set is accessible online.(14)

We downloaded the FASTQ data from SRA and processed using the same salmon/tximport pipeline we used for the astrocyte samples. The only difference was in the modeling step, where we converted the count data to log counts/million and estimated observation-level weights using the limma voom function. We also computed sample-specific weights that are intended to down-weight any samples that are not very similar to other samples of the same type. We then fit a conventional weighted linear model and made empirical Bayes adjusted contrasts between various groups. By incorporating sample-specific weights we were able to account for a single sample (LPS100 treated animal 4414T), which had significantly fewer reads, perhaps due to some technical problems.

*Venn diagrams*

We sought to learn which genes are unique to a given comparison, and which are shared between two or more of those comparisons. Thus, we generated Venn diagrams for three sets of comparisons. We made a Venn diagram for the three comparisons of treated vs. control, a two-way Venn diagram of the agonist + LPS versus either LPS alone or LPS plus antagonist (this shows that there is very little difference between LPS and LPS plus antagonist), and finally we made a Venn diagram of the one-hit versus two-hit for all three treatments. The genes in any intersection between comparisons had to be significantly differentially expressed in both (or all three, depending on the intersection) of the comparisons. In addition, the direction of change had to be the same as well. For example, if a given gene was differentially expressed in LPS100 vs control and LPS100+A10 vs control, and it is either up or down-regulated in both comparisons as well, then it was listed in the intersection between those two comparisons. If it was significant in both comparisons, but was up-regulated in one comparison, but down-regulated in the other, then it was listed in the unique portion of the Venn diagram for each comparison.

*Statistical analyses and data repository*

Generalized estimating equations (GEE) modeling approach was used to assess the effects of LPS and drug treatments on IL-1β. We used a linear scale response model with LPS/drug treatment group (main term "treatment") and presence or absence of second hit exposure (main term "hit") as predicting factors to assess their interactions using maximum likelihood estimate and Type III analysis with Wald Chi-square statistic. SPSS Version 25 was used for these analyses (IBM SPSS Statistics, IBM Corporation, Armonk, NY). Significance was assumed for $p < 0.05$. Results are provided as median {25-75} percentiles. Not all measurements were obtained for each animal studied, as indicated.

Ingenuity Pathway Analysis (IPA, Qiagen, 2019) was used for identification of signaling pathways unique to each treatment.

The raw RNAseq data has been deposited under GEO accession number [GSE123713](GSE123713). The analytical pipeline to allow reproduction of this analysis, in the form of an Rmarkdown document, will be made available upon request. Statistics from all comparisons (t-statistics, log fold changes, FDR values, dynamically linked Venn diagrams) can be found in supplemental document under the DOI [10.5281/zenodo.2609202](10.5281/zenodo.2609202).

## Results and Discussion

*Cytokine secretion profile*

The absolute values of IL-1β produced by naive (first hit) and second hit astrocytes at baseline, under LPS exposure and with preceding agonist or antagonist incubation are presented in Table 2. As reported, LPS treatment induced IL-1β increase in astrocytes.(14) There was no difference in first and second hit astrocytes at baseline or when exposed to LPS alone. Our focus here was on the effect of α7nAChR modulation on LPS-triggered IL-1β production. Consequently, we expressed the data as fold-changes compared to sham treatment (LPS, Fig. 2, LEFT). We present the response of primary microglia cultures side-by-side (Fig. 2, RIGHT) in form of re-analysis of previously published data.(3)

For astrocytes, the main terms "hit" and "treatment" as well as their interaction were significant (all $p<0.001$). Agonistic stimulation of the α7nAChR appeared, surprisingly, to result in a relative increase of IL-1β concentration, further potentiated in second hit astrocytes. This effect was absent in first hit astrocytes cultures treated with the α7nAChR antagonist and less pronounced but present in the second hit cultures.

For microglia, the main terms "hit" and "treatment" were not significant ($p=0.716$ and $p=0.666$, respectively), but their interaction was significant ($p=0.026$, cf. Fig. 1 in (3)).

The overall pattern of IL-1β levels in response to LPS exposure with prior α7nAChR stimulation was inverted in astrocytes compared to microglia. In astrocytes, both first hit and second hit cell cultures responded under antagonistic stimulation of α7nAChR with relative decrease of IL-1β production. In microglia, in contrast, first hit (naive) cultures behaved intuitively under the same conditions, showing a rise in IL-1β production. However, second hit microglia cultures, similar to second hit astrocytes, showed a drop in IL-1β production with α7nAChR antagonistic stimulation. In contrast again, agonistic α7nAChR stimulation in both second hit glia cultures resulted in higher than first hit IL-1β production, albeit, the magnitude of this rise was ~3.5-fold larger in astrocytes suggesting differences in sensitization of these glia cells to previous LPS exposure *in utero*. These counter-intuitive findings of α7nAChR stimulation in astrocytes on the individual IL-1β secretion stand in contrast to the transcriptome-level findings we discuss below. We attempt to tie together these results in the general discussion section.

*Whole transcriptome analysis*

Mapped reads aligned to any transcript from the Oar_v3.1 transcriptome at 66% which is good. Principal component analysis (PCA) showed large differences between the different treatment groups, and much smaller differences within each group, indicating that we have a good signal with likely many differentially expressed genes (DEG) to be found (Fig. 3).

We present in Table 3 the number of genes for each comparison as well as the top ten signaling pathways IPA identified. The results of the analysis on the entire data set are also accessible with search function via [this](#) repository.

PCA showed two transcriptome clusters (Fig. 3): control, LPS single-hit and second-hit astrocytes pre-treated with α7nAChR agonist and LPS single-hit and second-hit astrocytes pre-treated with α7nAChR antagonist. That is, a pro-inflammatory transcriptome astrocyte phenotype acquired *in vivo* or *in vitro* by LPS stimulation is reversed with α7nAChR agonistic stimulation. Conversely, antagonistic α7nAChR stimulation potentiates the pro-inflammatory astrocytic phenotype. The PCA level observations are substantiated further by the IPA analysis of key signaling pathways presented in Table 3. The visualization of the up or down regulation of the implicated pathways can be accessed on [GitHub](#) in its entirety. Here we focus on some key findings in top ten signaling pathways. LPS treatment triggered activation of pro-inflammatory signaling pathways NF-κB and neuroinflammation. Compared to LPS exposure alone, pretreatment with α7nAChR agonist reversed both signaling pathways activation. Conversely, pretreatment with α7nAChR antagonist up regulated these pro-inflammatory pathways. Albeit the pattern overall was similar to the effect of LPS alone, activation of these signaling pathways under α7nAChR blockade stood out: Toll-like receptor signaling and PI3K signaling. Direct comparison of α7nAChR agonistic and antagonistic stimulation yielded reduced activity of NF-κB and STAT3 pathways due to activation of α7nAChR, consistent with the expected intracellular anti-inflammatory effect of α7nAChR agonism. Another notable pathway activated in α7nAChR agonistically treated astrocytes was the Sirtuin signaling. Activation of this pathway in neurons and astrocytes has been implicated in AMPK-dependent neuroprotection from ischemic stroke.(20,21) Adenosine monophosphate kinase (AMPK) is a rapid key regulator of neuronal energy homeostasis implicated in fetal neuroinflammation.(22)

The second half of Table 3 documents some effects of astrocytes memory of LPS exposure *in vivo* when re-exposed *in vitro* (second hit effect). Notably, we found a perturbation of the iron homeostasis signaling pathway in second hit LPS treated astrocytes which persisted under pre-treatment with α7nAChR antagonist, but was reversed with α7nAChR agonist. Similar to our finding of second hit signature in microglia, here we [observed](#) hemoxygenase (HMOX)1 gene down regulation in second hit astrocytes compared to first hit cultures.(3,15) HMOX1 is a key gene of iron homeostasis. We observed a similar phenomenon in second hit fetal microglia compared to single hit microglia.(15)

Together, observations on Sirtuin and iron homeostasis signaling reinforce the previously reported dual role of energy metabolism in determining inflammatory phenotype in glia cells.(3,15)

Based on the IPA analysis within the [Venn diagrams](#), the top down-regulated signaling pathway unique to LPS treatment was Ephrin A signaling (log(p-value) 3.21, Table 5). Analysis of genes unique to α7nAChR agonist treatment showed a reduction of this down-regulation (log(p-value)1.58, Table 5). Analysis of genes unique to α7nAChR antagonist treatment had no significant effect on this signaling pathway (Table 5). Ephrin signalling has been implicated in neuroprotective astrocyte phenotype.(23) Our findings suggest a neuroprotective effect of α7nAChR agonism on ephrin signaling pathway.

Consistent with the notion of cholinergic signaling involved in stress axis regulation (3), POMC was the second (by IPA ranking) highest up-regulated gene under cholinergic agonist treatment, with a log ratio of 3.388 (Cf. [GitHub repository](#)).

*Comparison to fetal sheep microglial transcriptome*

PCA in Fig. 4 (top) shows that the main differences between astrocytes and microglia RNAseq data are captured on the first two principal components, so a 2D plot may be more useful (Fig. 4, bottom). The intra-group variability is smaller for the astrocytes compared to the microglia. This may have to do with the total library sizes: there were several microglia samples with very few reads that aligned to any known transcript.

We compared the IPA-identified top signaling pathways in microglia and astrocytes under the LPS and α7nAChR signaling manipulation. Table 5 presents the number of genes for each comparison and the corresponding findings of the IPA signaling pathway analysis. Overall, the response patterns to LPS and modulation of α7nAChR signaling were similar between the two glia cell types. The signaling pathways common to both astrocytes and microglia are bolded in Tables 3 and 4. Intuitively, common pathways activated due to LPS included neuroinflammation signaling and NF-κB signaling in some, but not all comparisons. However, overall, the overlap on the level of signaling pathway was rather minimal which may explain the strong separation by cell type on PCA. It is remarkable that astrocytes, not microglia - the primary immune cells of the brain, were characterized by unique inhibition patterns of STAT3 pathway due to agonistic stimulation of α7nAChR prior to LPS exposure.

The presented astrocytes - microglia comparison has limitations. The differences between the two cell types may be exaggerated by the inevitable technical differences (e.g., reagents). However, all these experiments were run from the same cohort, same animals (in some cases), at adjacent times and by the same people.

*Do fetal neuroinflammation and stress mediate an increased risk for autism spectrum disorder?*

LPS effects on key genes involved in stress axis activity raised the question about the poorly understood role of astrocytes in the signaling pathways of neuroinflammation and prenatal stress (PS).(24)

Indeed, PS is accompanied by inflammation in the mother and offspring.(24–28) (29) Both, PS and fetal neuroinflammation have been implicated in the etiology of ASD.(30,31) PS increases expression of glutamate (Glu) transporter vGluT1 (SLC17A7) resulting in higher levels of GLT1.(32,33) (33)

Here, we sought to verify if the exposure of fetal astrocytes to LPS induces upregulation of Glu transporters in glia akin to PS.(32,33)

Given the known relationship between PS and VGLUT1 expression, we used IPA to annotate VGLUT1 gene network with our findings to test for evidence that the network is being perturbed by LPS treatment. Across all treatment comparisons, two DEGs were identified in astrocytes (common to all comparisons): JAK2 (2.732 (log ratio) upregulated, FDR 3.91E-05) and SLC1A2 (2.350 upregulated, FDR 1.80E-04). JAK1 signaling is involved in glucocorticoid receptor signaling(34) and JAK1/2 signaling is involved in iron homeostasis signaling pathways(35), whereas JAK/Stat is involved in IL-6 signaling pathways. SLC1A2 is also known as GLT-1 or glial high affinity glutamate transporter; it is implicated in glutamate receptor signaling and neuroinflammation signaling pathways.(36) Upregulation of glial GLT-1 in the hippocampus has been reported after chronic stress due to its control by glucocorticoids.(37–39)

Again assessing the present findings together with the previously published RNAseq data from the identical experiment in ovine fetal microglia (3), we found GLT-1 to be upregulated in microglia and astrocytes regardless cholinergic manipulation (for details see [GitHub repository](#)). Albeit both pure primary cultures were exposed to LPS, we found the up regulation of JAK2 to be unique to astrocytes. This finding is conceptually in line with studies reporting brain cell-specific signalling pathways behavior in response to IL-1β.(40) Much remains to be learned about the differences between second messenger signaling cascades involved in astrocytes, microglia and neurons in the developing fetal brain exposed to inflammatory stimuli. Furthermore, there is still a paucity of data about differences across the species for these signaling pathways.

*General discussion*

Fetal sheep is the classic model of fetal physiology and neuroscience.(41) It has been used successfully for both integrative physiological as well as genomic studies. In the present study, we expand the recently published series of experiments in the same animal model using primary microglia cultures to now include primary astrocytes cultures.(3,14–16)

Acetylcholine is synthesized by cultured microglia and astrocyte in mouse and rat (42,43), however, there is no information from other species. It is plausible that the fetal sheep brain astrocytes in culture produce acetylcholine, because they appear to respond to α7nAChR stimulation. However, we do not know how much acetylcholine is produced, and whether or not the speculated endogenous acetylcholine can activate the α7nAChR. This remains worthwhile to investigate in future studies.

The antagonist drug for α7nAChR we used, α-bungarotoxin, is a selective inhibitor for α7 receptors acting by preventing the opening of nicotinic receptor-associated ion channels (Tocris α-bungarotoxin datasheet). By using optimized dose, we treated our astrocyte cells with α-bungarotoxin one hour prior to LPS exposure, which would block LPS-induced cytokine production in the cells.

LPS exposure had the anticipated effect of increasing IL-1β production in astrocytes and we observed this consistently on the transcriptome level. However, our findings on protein level following pre-incubation with α7nAChR antagonist or agonist do not align with those on the transcriptome level: present results on protein level do not show a clear hypothesized effect of α7nAChR agonism or antagonism on LPS secretion in ovine fetal astrocytes. Such discordant behavior on protein and transcriptome levels has been reported and studied systematically to represent the rule rather than an exception to cellular biology in general.(44)

Future studies will need to explore the protein responses in more depth and in different species to further delineate astrocytic behavior under α7nAChR stimulation, especially the peculiarly opposite effect of endotoxin memory in astrocytes and microglia on IL-1β secretion.

Microglia - astrocyte ensemble interactions need to be studied to bridge the methodological gap between *in vitro* experimental design and the *in situ* physiology. This can be done in co-cultures, feasible in ovine species, for example.(14) Recent study in mice highlighted the importance of microglia - astrocyte interactions for understanding the polarization dynamics of astrocytes.(45)

In an adult rodent model, cholinergic signalling reduced stress responsiveness via CRH receptor 1 with positive behavioural changes.(46) We identified POMC as up regulated under α7nAChR stimulation. Considering that CRH is an upstream regulator of POMC, another question for future studies is whether we can implicate a direct interaction with the CRH receptor 1 in the developing brain.

We found that within its interaction network, GLT-1 was upregulated in microglia and astrocytes regardless of cholinergic manipulation, while the up-regulation of JAK1/2 was unique to astrocytes. This is in line with studies showing brain region specific overexpression of vGluT-1 (SLC17A7) both due to endotoxin stress and due to PS and puts the LPS exposure in the context of a more general brain stress exposure paradigm presenting with shared response patterns of neuroinflammation and metabolic adaptations in astrocytes.(13,47,48) PS has long lasting consequences on α7nACh-ergic signaling in frontal cortex and hippocampus reducing the expression of α7nAChR protein expression in the brain of adult rats.(47)

It is plausible to conclude that low grade neuroinflammation results in changes similar to those induced by PS with regard to reprogramming astrocytes to a higher glutamate uptake. As shown elsewhere (49), PS and exposure to endotoxin may act synergistically to exacerbate the impairment of neuron-glial glutamatergic interaction. For endotoxin exposure, this process is not subject to cholinergic modulation. Whether or not PS effects alone on astrocytes glutamate uptake can be ameliorated or reversed by α7nAChR agonism remains to be investigated.

The phylogenetically conserved interaction between neuroinflammation and chronic stress has been the subject of multiple studies (24), yet we are only beginning to unravel the complex web of interactions, across developmental stages, organs, cell types and species-specific differences, which connect these two phenomena. While the role of microglia in this context has been appreciated, the response of astrocytes we report here and their behavior on protein level under second hit scenario are novel observations warranting further studies in different species.

In summary, we show that genes involved in stress memory of the offspring are also impacted by LPS stress, this impact is further altered by a second hit (memory) and that such memory of LPS stress is amenable to cholinergic treatment via α7nAChR. It remains to be validated in future studies whether, when and which stimulation of the α7nAChR is favourable.

**AUTHOR CONTRIBUTIONS**

MCa, PB, GF, AD, and MGF are responsible for the conception and design.
MCa, MCo, HLL, MW, LDD, PB, GF, AD and MGF acquired data.
MCo, MCa, HLL, MW, JWM, MD, TKB, JS and MGF did the analysis and interpretation of data.
MCo, MCa, JWM, TKB and MGF drafted the manuscript.
MCo, MCa, MW, JWM, JS, MCA, TKB and MGF are responsible for revising it for intellectual content.
MCo, MCa, HLL, MW, LDD, JWM, PB, GF, AD, JS, MCA, TKB, MD and MGF gave final approval of the completed manuscript.

**Funding**: Supported by grants from the Canadian Institute of Health Research (CIHR) (MGF); Fonds de la recherche en santé du Québec (FRSQ) (MGF) and Molly Towell Perinatal Research Foundation (MGF); QTNPR (by CIHR) (LDD). Supported in part by Illumina Inc.

**ACKNOWLEDGEMENTS**
The authors thank Dr. Jack Antel lab, especially Manon Blain, and Dr. Craig Moore for invaluable assistance with the cell culture protocol, St-Hyacinthe CHUV team for technical assistance and Jan Hamanishi for graphical design. We also thank Dr. Michael Dorschner lab for skilful assistance with RNA samples pipeline on the Illumina platform and Lu Wang for assistance in RNAseq bioinformatics pipeline.

**Figure legends**

**Figure 1. Experimental design of modulating glial α7nAChR signaling in a double-hit fetal sheep model.** *In vivo, in vitro* and RNAseq experiments are illustrated. *In vivo* study includes Control (saline) or LPS-exposed sheep fetuses. LPS was administered intravenously to the instrumented fetus at 400 ng/fetus/day for two consecutive days 24 hours apart, so called first hit, inducing fetal inflammatory response with rising IL-6, but without cardiovascular component. For the *in vitro* study, cultured cells (microglia or astrocytes) were derived from an *in vivo* Control animal, named Naïve, or from an *in vivo* LPS-exposed animal, named second hit (SH). There weare 8 experimental groups: naïve Control (NC, vehicle *in vivo*, vehicle *in vitro*), naïve LPS (NL, sham in reference to α7nAChR manipulation), naïve exposed to α-Bungarotoxin (NB, i.e., α7nAChR inhibition, preincubated followed by LPS exposure), naïve exposed to AR-R17779 (NA, i.e., α7nAChR stimulation, preincubated followed by LPS exposure), and each respective second-hit groups (SH, LPS *in vivo*). Reproduced with permission.(16)

**Figure 2. LEFT.** IL-1β secretion in ovine primary astrocyte cultures in response to six hours LPS exposure without or with pre-incubation with α7nAChR antagonist (B100) or agonist (A10) for one hour. Single hit, *in vitro* only LPS exposure; second hit, *in vivo* systemic and subsequent *in vitro* LPS exposure four to five weeks later. Y axis shows fold changes in IL-1β in relation to baseline secretion levels on log scale. Generalized estimating equations (GEE) modeling results are presented in text and no significance marks are provided in the figure. Briefly, we found significant main term effects (p=0.019) "treatment" (LPS and α7nAChR drug) as well as main term "hit" (p=0.010), *i.e.*, the contribution of *in vivo* LPS exposure, the second hit effect on the IL-1β secretion profile. Results are provided as median {25-75} percentiles. **RIGHT**. Identical experimental results from microglia studies are presented for comparison. The main terms "hit" and "treatment" were not significant (p=0.716 and p=0.666, respectively), but their interaction was significant (p=0.026, cf. Fig. 1 in (3) where the original results have been published.

**Figure 3.** Static 3D plot of the astrocyte RNA-Seq data with single and double-hit LPS treatment. The angle of the plot was chosen to give the best viewpoint to show differences between the sample types. Note that controls and α7nAChR agonistically pre-treated astrocytes cluster together and separately from those exposed to LPS w/o or with antagonistic α7nAChR pre-treatment. Note here that there are three of the LPS_100_1_hit samples in the plot shown in Fig. 3; it just so happens that the third sample is obscured by the uppermost LPS+B100_1_hit sample.

**Figure 4. TOP.** PCA plot of microglia and astrocyte samples. Here we can see that astrocytes and microglia separate on the first principal component, and the second principal component captures the LPS treatment differences. The third principal component captures some intra-treatment variability for the microglia samples, particularly for one of the LPS treated microglia samples. **BOTTOM.** PCA plot of microglia and astrocyte samples, showing just the first two principal components. The largest differences appear to be between the cell types.

**References**

1. Jassam YN, Izzy S, Whalen M, McGavern DB, El Khoury J. Neuroimmunology of Traumatic Brain Injury: Time for a Paradigm Shift. *Neuron* (2017) **95**:1246–1265. doi:10.1016/j.neuron.2017.07.010

2. Kalkman HO, Feuerbach D. Modulatory effects of α7 nAChRs on the immune system and its relevance for CNS disorders. *Cell Mol Life Sci* (2016) **73**:2511–2530. doi:10.1007/s00018-016-2175-4

3. Cortes M, Cao M, Liu HL, Moore CS, Durosier LD, Burns P, Fecteau G, Desrochers A, Barreiro LB, Antel JP, et al. α7 nicotinic acetylcholine receptor signaling modulates the inflammatory phenotype of fetal brain microglia: first evidence of interference by iron homeostasis. *Sci Rep* (2017) **7**:10645. doi:10.1038/s41598-017-09439-z

4. Kiguchi N, Kobayashi D, Saika F, Matsuzaki S, Kishioka S. Inhibition of peripheral macrophages by nicotinic acetylcholine receptor agonists suppresses spinal microglial activation and neuropathic pain in mice with peripheral nerve injury. *J Neuroinflammation* (2018) **15**:96. doi:10.1186/s12974-018-1133-5

5. Frasch MG, Szynkaruk M, Prout AP, Nygard K, Cao M, Veldhuizen R, Hammond R, Richardson BS. Decreased neuroinflammation correlates to higher vagus nerve activity fluctuations in near-term ovine fetuses: a case for the afferent cholinergic anti-inflammatory pathway? *J Neuroinflammation* (2016) **13**:103. doi:10.1186/s12974-016-0567-x

6. Han Z, Li L, Wang L, Degos V, Maze M, Su H. Alpha-7 nicotinic acetylcholine receptor agonist treatment reduces neuroinflammation, oxidative stress, and brain injury in mice with ischemic stroke and bone fracture. *J Neurochem* (2014) **131**:498–508. Available at: https://onlinelibrary.wiley.com/doi/abs/10.1111/jnc.12817

7. de Jonge WJ, Ulloa L. The alpha7 nicotinic acetylcholine receptor as a pharmacological target for inflammation. *Br J Pharmacol* (2007) **151**:915–929. doi:10.1038/sj.bjp.0707264

8. Shytle RD, Mori T, Townsend K, Vendrame M, Sun N, Zeng J, Ehrhart J, Silver AA, Sanberg PR, Tan J. Cholinergic modulation of microglial activation by alpha 7 nicotinic receptors. *J Neurochem* (2004) **89**:337–343. doi:10.1046/j.1471-4159.2004.02347.x

9. Rosas-Ballina M, Olofsson PS, Ochani M, Valdes-Ferrer SI, Levine YA, Reardon C, Tusche MW, Pavlov VA, Andersson U, Chavan S, et al. Acetylcholine-synthesizing T cells relay neural signals in a vagus nerve circuit. *Science* (2011) **334**:98–101. doi:10.1126/science.1209985

10. Cheyuo C, Jacob A, Wu R, Zhou M, Coppa GF, Wang P. The parasympathetic nervous system in the quest for stroke therapeutics. *J Cereb Blood Flow Metab* (2011) **31**:1187–1195. doi:10.1038/jcbfm.2011.24

11. Cao J, Lu K-H, Powley TL, Liu Z. Vagal nerve stimulation triggers widespread responses and alters large-scale functional connectivity in the rat brain. *PLoS One* (2017) **12**:e0189518. doi:10.1371/journal.pone.0189518

12. al-Haddad BJS, Jacobsson B, Chabra S, Modzelewska D, Olson EM, Bernier R, Enquobahrie DA, Hagberg H, Östling S, Rajagopal L, et al. Long-term Risk of Neuropsychiatric Disease After Exposure to Infection In Utero. *JAMA Psychiatry* (2019) doi:10.1001/jamapsychiatry.2019.0029

13. Elovitz MA, Brown AG, Breen K, Anton L, Maubert M, Burd I. Intrauterine inflammation, insufficient to induce parturition, still evokes fetal and neonatal brain injury. *Int J Dev Neurosci* (2011) **29**:663–671. doi:10.1016/j.ijdevneu.2011.02.011

14. Cortes M, Cao M, Liu HL, Burns P, Moore C, Fecteau G, Desrochers A, Barreiro LB, Antel JP, Frasch MG. RNAseq profiling of primary microglia and astrocyte cultures in near-term ovine fetus: A glial in vivo-in vitro multi-hit paradigm in large mammalian brain. *J Neurosci Methods* (2017) **276**:23–32. doi:10.1016/j.jneumeth.2016.11.008

15. Cao M, Cortes M, Moore CS, Leong SY, Durosier LD, Burns P, Fecteau G, Desrochers A, Auer RN, Barreiro LB, et al. Fetal microglial phenotype in vitro carries memory of prior in vivo exposure to inflammation. *Front Cell Neurosci* (2015) **9**:294. doi:10.3389/fncel.2015.00294

16. Frasch MG, Burns P, Benito J, Cortes M, Cao M, Fecteau G, Desrochers A. Sculpting the Sculptors: Methods for Studying the Fetal Cholinergic Signaling on Systems and Cellular Scales. *Methods Mol Biol* (2018) **1781**:341–352. doi:10.1007/978-1-4939-7828-1_18

17. Patro R, Duggal G, Kingsford C. Salmon: Accurate, Versatile and Ultrafast Quantification from RNA-seq Data using Lightweight-Alignment. *bioRxiv* (2015)021592. doi:10.1101/021592

18. Soneson C, Love MI, Robinson MD. Differential analyses for RNA-seq: transcript-level estimates improve gene-level inferences. *F1000Res* (2015) **4**:1521. doi:10.12688/f1000research.7563.2

19. Smyth GK. Linear models and empirical bayes methods for assessing differential expression in microarray experiments. *Stat Appl Genet Mol Biol* (2004) **3**:Article3. doi:10.2202/1544-6115.1027

20. Wang P, Xu T-Y, Guan Y-F, Tian W-W, Viollet B, Rui Y-C, Zhai Q-W, Su D-F, Miao C-Y. Nicotinamide phosphoribosyltransferase protects against ischemic stroke through SIRT1-dependent adenosine monophosphate-activated kinase pathway. *Ann Neurol* (2011) **69**:360–374. doi:10.1002/ana.22236

21. Li D, Liu N, Zhao H-H, Zhang X, Kawano H, Liu L, Zhao L, Li H-P. Interactions between Sirt1 and MAPKs regulate astrocyte activation induced by brain injury in vitro and in vivo. *J Neuroinflammation* (2017) **14**:67. doi:10.1186/s12974-017-0841-6

22. Frasch MG. Putative Role of AMPK in Fetal Adaptive Brain Shut-Down: Linking Metabolism and Inflammation in the Brain. *Front Neurol* (2014) **5**:150. doi:10.3389/fneur.2014.00150

23. Tyzack GE, Hall CE, Sibley CR, Cymes T, Forostyak S, Carlino G, Meyer IF, Schiavo G, Zhang S-C, Gibbons GM, et al. A neuroprotective astrocyte state is induced by neuronal signal EphB1 but fails in ALS models. *Nat Commun* (2017) **8**:1164.

doi:10.1038/s41467-017-01283-z

24. Deak T, Kudinova A, Lovelock DF, Gibb BE, Hennessy MB. A multispecies approach for understanding neuroimmune mechanisms of stress. *Dialogues Clin Neurosci* (2017) **19**:37–53. Available at: https://www.ncbi.nlm.nih.gov/pubmed/28566946

25. Diz-Chaves Y, Astiz M, Bellini MJ, Garcia-Segura LM. Prenatal stress increases the expression of proinflammatory cytokines and exacerbates the inflammatory response to LPS in the hippocampal formation of adult male mice. *Brain Behav Immun* (2013) **28**:196–206. doi:10.1016/j.bbi.2012.11.013

26. Bronson SL, Bale TL. Prenatal stress-induced increases in placental inflammation and offspring hyperactivity are male-specific and ameliorated by maternal antiinflammatory treatment. *Endocrinology* (2014) **155**:2635–2646. doi:10.1210/en.2014-1040

27. Wu S, Gennings C, Wright RJ, Wilson A, Burris HH, Just AC, Braun JM, Svensson K, Zhong J, Brennan KJM, et al. Prenatal Stress, Methylation in Inflammation-Related Genes, and Adiposity Measures in Early Childhood: the Programming Research in Obesity, Growth Environment and Social Stress Cohort Study. *Psychosom Med* (2018) **80**:34–41. doi:10.1097/PSY.0000000000000517

28. Shapiro GD, Fraser WD, Frasch MG, Séguin JR. Psychosocial stress in pregnancy and preterm birth: associations and mechanisms. *J Perinat Med* (2013) **41**:631–645. doi:DOI 10.1515/jpm-2012-0295

29. Christian LM, Franco A, Glaser R, Iams JD. Depressive symptoms are associated with elevated serum proinflammatory cytokines among pregnant women. *Brain Behav Immun* (2009) **23**:750–754. doi:10.1016/j.bbi.2009.02.012

30. El-Ansary A, Al-Ayadhi L. Neuroinflammation in autism spectrum disorders. *J Neuroinflammation* (2012) **9**:265. doi:10.1186/1742-2094-9-265

31. Kinney DK, Munir KM, Crowley DJ, Miller AM. Prenatal stress and risk for autism. *Neurosci Biobehav Rev* (2008) **32**:1519–1532. doi:10.1016/j.neubiorev.2008.06.004

32. Barros VG, Duhalde-Vega M, Caltana L, Brusco A, Antonelli MC. Astrocyte-neuron vulnerability to prenatal stress in the adult rat brain. *J Neurosci Res* (2006) **83**:787–800. doi:10.1002/jnr.20758

33. Barros VG, Berger MA, Martijena ID, Sarchi MI, Perez AA, Molina VA, Tarazi FI, Antonelli MC. Early adoption modifies the effects of prenatal stress on dopamine and glutamate receptors in adult rat brain. *J Neurosci Res* (2004) **76**:488–496. doi:10.1002/jnr.20119

34. Pace TWW, Miller AH. Cytokines and glucocorticoid receptor signaling. Relevance to major depression. *Ann N Y Acad Sci* (2009) **1179**:86–105. doi:10.1111/j.1749-6632.2009.04984.x

35. Schmidt PJ. Regulation of Iron Metabolism by Hepcidin under Conditions of Inflammation. *J Biol Chem* (2015) **290**:18975–18983. doi:10.1074/jbc.R115.650150

36. Jauregui-Huerta F, Ruvalcaba-Delgadillo Y, Gonzalez-Castañeda R, Garcia-Estrada J,

Gonzalez-Perez O, Luquin S. Responses of glial cells to stress and glucocorticoids. *Curr Immunol Rev* (2010) **6**:195–204. doi:10.2174/157339510791823790

37. Autry AE, Grillo CA, Piroli GG, Rothstein JD, McEwen BS, Reagan LP. Glucocorticoid regulation of GLT-1 glutamate transporter isoform expression in the rat hippocampus. *Neuroendocrinology* (2006) **83**:371–379. doi:10.1159/000096092

38. Reagan LP, Rosell DR, Wood GE, Spedding M, Muñoz C, Rothstein J, McEwen BS. Chronic restraint stress up-regulates GLT-1 mRNA and protein expression in the rat hippocampus: reversal by tianeptine. *Proc Natl Acad Sci U S A* (2004) **101**:2179–2184. doi:10.1073/pnas.0307294101

39. Zschocke J, Bayatti N, Clement AM, Witan H, Figiel M, Engele J, Behl C. Differential promotion of glutamate transporter expression and function by glucocorticoids in astrocytes from various brain regions. *J Biol Chem* (2005) **280**:34924–34932. doi:10.1074/jbc.M502581200

40. Srinivasan D, Yen J-H, Joseph DJ, Friedman W. Cell Type-Specific Interleukin-1β Signaling in the CNS. *J Neurosci* (2004) **24**:6482–6488. doi:10.1523/JNEUROSCI.5712-03.2004

41. Morrison JL, Berry MJ, Botting KJ, Darby JRT, Frasch MG, Gatford KL, Giussani DA, Gray CL, Harding R, Herrera EA, et al. Improving pregnancy outcomes in humans through studies in sheep. *Am J Physiol Regul Integr Comp Physiol* (2018) doi:10.1152/ajpregu.00391.2017

42. Wessler I, Kirkpatrick CJ, Racké K. The cholinergic "pitfall": acetylcholine, a universal cell molecule in biological systems, including humans. *Clin Exp Pharmacol Physiol* (1999) **26**:198–205. Available at: https://www.ncbi.nlm.nih.gov/pubmed/10081614

43. Wessler I, Reinheimer T, Klapproth H, Schneider FJ, Racké K, Hammer R. Mammalian glial cells in culture synthesize acetylcholine. *Naunyn Schmiedebergs Arch Pharmacol* (1997) **356**:694–697. Available at: https://www.ncbi.nlm.nih.gov/pubmed/9402051

44. Ghazalpour A, Bennett B, Petyuk VA, Orozco L, Hagopian R, Mungrue IN, Farber CR, Sinsheimer J, Kang HM, Furlotte N, et al. Comparative analysis of proteome and transcriptome variation in mouse. *PLoS Genet* (2011) **7**:e1001393. doi:10.1371/journal.pgen.1001393

45. Rothhammer V, Borucki DM, Tjon EC, Takenaka MC, Chao C-C, Ardura-Fabregat A, de Lima KA, Gutiérrez-Vázquez C, Hewson P, Staszewski O, et al. Microglial control of astrocytes in response to microbial metabolites. *Nature* (2018) **557**:724–728. doi:10.1038/s41586-018-0119-x

46. Farrokhi CB, Tovote P, Blanchard RJ, Blanchard DC, Litvin Y, Spiess J. Cortagine: behavioral and autonomic function of the selective CRF receptor subtype 1 agonist. *CNS Drug Rev* (2007) **13**:423–443. doi:10.1111/j.1527-3458.2007.00027.x

47. Baier CJ, Pallares ME, Adrover E, Monteleone MC, Brocco MA, Barrantes FJ, Antonelli MC. Prenatal restraint stress decreases the expression of alpha-7 nicotinic receptor in the brain of adult rat offspring. *Stress* (2015) **18**:435–445.

doi:10.3109/10253890.2015.1022148

48. Adrover E, Pallarés ME, Baier CJ, Monteleone MC, Giuliani FA, Waagepetersen HS, Brocco MA, Cabrera R, Sonnewald U, Schousboe A, et al. Glutamate neurotransmission is affected in prenatally stressed offspring. *Neurochem Int* (2015) **88**:73–87. doi:10.1016/j.neuint.2015.05.005

49. Cumberland AL, Palliser HK, Rani P, Walker DW, Hirst JJ. Effects of combined IUGR and prenatal stress on the development of the hippocampus in a fetal guinea pig model. *J Dev Orig Health Dis* (2017) **8**:584–596. doi:10.1017/s2040174417000307

**Table 1. Sample inventory for RNAseq study in astrocytes samples.**

| Biological replicates | Serial# | Sample | Treatment | LPS hit number |
|---|---|---|---|---|
| Naïve R1 | 1 | 4414T1 | Ctrl | 0 |
|  | 2 | 4414T1 | LPS100 | 1 |
|  | 3 | 4414T1 | LPS100+B100 | 1 |
|  | 4 | 4414T1 | LPS100+A10 | 1 |
| Naïve R2 | 5 | 4414T2 | Ctrl | 0 |
|  | 6 | 4414T2 | LPS100 | 1 |
|  | 7 | 4414T2 | LPS100+B100 | 1 |
|  | 8 | 4414T2 | LPS100+A10 | 1 |
| Naïve R3 | 9 | 4502P | Ctrl | 0 |
|  | 10 | 4502P | LPS100 | 1 |
|  | 11 | 4502P | LPS100+B100 | 1 |
|  | 12 | 4502P | LPS100+A10 | 1 |
| Second hit R1 | 13 | 711P | Ctrl | 1 |
|  | 14 | 711P | LPS100 | 2 |
|  | 15 | 711P | LPS100+B100 | 2 |

Four replicates were used for RNAseq, with one control and three treatment groups (Ctrl, LPS100, LPS100+A10, LPS100+B100) per replicate. There were 3 replicates of naïve, one replicate of second-hit, with one treatment missing in this second-hit replicate (second-hit + Agonist), fifteen samples in total were assessed with RNAseq in this study. The instrumented fetuses were designated as primary fetus, identified as animal ID+P (stands for primary), whereas animal ID+T stands for non-instrumented twins.

**Table 2.** Astrocytes IL-1β secretion expressed as absolute values* in pg/ml (median and 25-75%).

| LPS exposure | Ctrl | LPS | B100 | A10 |
|---|---|---|---|---|
| **Single hit** | 1 (1,25) | 429 (358,1034) | 336 (285,928) | 750 (439,1495) |
| **Second hit** | 1 (1,1) | 16 (15,16) | 20 (17, 22) | 99 (95, 103) |

* Values set to 1 where no signal was detected by the cytokine assay to compute fold-changes for Fig. 2.

**Table 3. Differentially expressed genes from the fetal sheep astrocytes whole transcriptome analysis: naïve and second hit astrocytes after modulation of α7nAChR signaling.** Differential analysis of count data was done with the Bioconductor limma package. Differentially expressed genes were selected, based on a 1.5-fold change and an FDR < 0.05. Up regulation and down regulation represent positive and negative log2 fold changes, respectively. For details on "raw gene" level, see our GitHub repository or directly here. Bold font highlights pathways common with microglia. Orange: positive z-score; blue: negative z-score. For further details as raw data and visualized activity patterns see GitHub.

| Comparison | # genes | Top 10 signaling pathways | -log(p-value) |
|---|---|---|---|
| Single hit: LPS100 vs Control | 1835 | NF-κB Signaling | 15.8 |
| | | Role of Pattern Recognition Receptors in Recognition of Bacteria & Viruses | 14 |
| | | **Role of Macrophages, Fibroblasts and Endothelial Cells in Rheumatoid Arthritis** | 13.7 |
| | | **Dendritic Cell Maturation** | 13 |
| | | **Neuroinflammation Signaling Pathway** | 12.8 |
| | | **Role of IL-17A in Arthritis** | 12.2 |
| | | Activation of IRF by Cytosolic Pattern Recognition Receptors | 11.6 |
| | | Death Receptor Signaling | 11.3 |
| | | Th1 and Th2 Activation Pathway | 10.7 |
| | | TREM1 Signaling | 10.5 |
| Single hit: LPS100+A10 vs Control | 1725 | NF-κB Signaling | 14.7 |
| | | **Role of Macrophages, Fibroblasts and Endothelial Cells in Rheumatoid Arthritis** | 13.7 |
| | | Role of Osteoblasts, Osteoclasts and Chondrocytes in Rheumatoid Arthritis | 13.1 |
| | | Granulocyte Adhesion and Diapedesis | 12.7 |
| | | Role of Pattern Recognition Receptors in Recognition of Bacteria and Viruses | 12.5 |
| | | Hepatic Cholestasis | 12.5 |
| | | Toll-like Receptor Signaling | 11.7 |
| | | Axonal Guidance Signaling | 11.5 |
| | | IL-10 Signaling | 11.1 |
| | | Neuroinflammation Signaling Pathway | 10.7 |
| Single hit: LPS100+B100 vs Control | 1744 | NF-κB Signaling | 16.8 |
| | | **Role of Macrophages, Fibroblasts and Endothelial Cells in Rheumatoid Arthritis** | 16 |
| | | Dendritic Cell Maturation | 15.9 |
| | | Neuroinflammation Signaling Pathway | 15 |
| | | Role of Pattern Recognition Receptors in Recognition of Bacteria & Viruses | 14.8 |
| | | TREM1 Signaling | 12.8 |
| | | Role of IL-17A in Arthritis | 12.7 |
| | | T Cell Exhaustion Signaling Pathway | 11.9 |
| | | **Toll-like Receptor Signaling** | 11.8 |
| | | **PI3K Signaling in B Lymphocytes** | 11.6 |
| Single hit: LPS100+A10 vs LPS100 | 273 | Granulocyte Adhesion and Diapedesis | 6.23 |
| | | Pathogenesis of Multiple Sclerosis | 5.81 |
| | | Agranulocyte Adhesion and Diapedesis | 5.12 |
| | | Th1 and Th2 Activation Pathway | 4.45 |
| | | Th2 Pathway | 4.4 |
| | | LPS/IL-1 Mediated Inhibition of RXR Function | 3.81 |
| | | NF-κB Signaling | 3.71 |
| | | Role of Osteoblasts, Osteoclasts and Chondrocytes in Rheumatoid Arthritis | 3.63 |
| | | Inhibition of Angiogenesis by TSP1 | 3.33 |
| | | STAT3 Pathway | 3.25 |
| Single hit: LPS100+B100 vs LPS100 | 0 | | |
| Single hit: LPS100+A10 vs LPS100+B100 | 292 | Granulocyte Adhesion and Diapedesis | 5.25 |
| | | Agranulocyte Adhesion and Diapedesis | 4.99 |
| | | LPS/IL-1 Mediated Inhibition of RXR Function | 4.38 |
| | | Hepatic Fibrosis / Hepatic Stellate Cell Activation | 4.35 |
| | | NF-κB Signaling | 4.33 |
| | | Role of Osteoblasts, Osteoclasts and Chondrocytes in Rheumatoid Arthritis | 4.18 |
| | | Pathogenesis of Multiple Sclerosis | 3.97 |
| | | STAT3 Pathway | 3.94 |
| | | p53 Signaling | 3.57 |
| | | Sirtuin Signaling Pathway | 3.36 |
| LPS100: single hit vs second hit | 3761 | Hepatic Fibrosis / Hepatic Stellate Cell Activation | 10.7 |
| | | Fcγ Receptor-mediated Phagocytosis in Macrophages and Monocytes | 7.69 |
| | | Leukocyte Extravasation Signaling | 6.82 |

| | | Signaling by Rho Family GTPases | 6.16 |
| | | Iron homeostasis signaling pathway | 6.02 |
| | | LXR/RXR Activation | 5.36 |
| | | Epithelial Adherens Junction Signaling | 4.99 |
| | | Role of Osteoblasts, Osteoclasts and Chondrocytes in Rheumatoid Arthritis | 4.96 |
| | | Axonal Guidance Signaling | 4.91 |
| | | Tec Kinase Signaling | 4.82 |
| LPS100+A10: single hit vs second hit | 3307 | Hepatic Fibrosis / Hepatic Stellate Cell Activation | 13 |
| | | Leukocyte Extravasation Signaling | 8.79 |
| | | Neuroinflammation Signaling Pathway | 7.18 |
| | | Fcγ Receptor-mediated Phagocytosis in Macrophages and Monocytes | 6.99 |
| | | Axonal Guidance Signaling | 6.93 |
| | | Phagosome Formation | 6.85 |
| | | Role of Pattern Recognition Receptors in Recognition of Bacteria and Viruses | 6.67 |
| | | Agranulocyte Adhesion and Diapedesis | 6.24 |
| | | GP6 Signaling Pathway | 6.1 |
| | | Granulocyte Adhesion and Diapedesis | 5.92 |
| LPS100+B100: single hit vs second hit | 3860 | Hepatic Fibrosis / Hepatic Stellate Cell Activation | 8.89 |
| | | Fcγ Receptor-mediated Phagocytosis in Macrophages and Monocytes | 8.03 |
| | | Leukocyte Extravasation Signaling | 7.26 |
| | | Epithelial Adherens Junction Signaling | |

| | | Iron homeostasis signaling pathway | 5.94 |
| --- | --- | --- | --- |
| | | Signaling by Rho Family GTPases | 5.78 |
| | | Axonal Guidance Signaling | 5.51 |
| | | Endothelin-1 Signaling | 5.18 |
| | | Role of Macrophages, Fibroblasts and Endothelial Cells in Rheumatoid Arthritis | 5.04 |
| | | | 5.01 |
| | | Phagosome Formation | 4.99 |

Significant genes in astrocytes cultures at an FDR < 0.05 and a 1.5-fold change sorted on log(p-value).

Table 4. Differentially expressed genes from the fetal sheep microglia whole transcriptomes analysis. Differential analysis of count data was done with the Bioconductor limma package. Differentially expressed genes were selected, based on a 1.5-fold change and an FDR < 0.05. Up regulation and down regulation represent positive and negative log2 fold changes, respectively. For details on "raw gene" level, see our GitHub repository or directly here. Bold font highlights signaling pathways common with astrocytes. Orange: positive z-score; blue: negative z-score. For further details as raw data and visualized activity patterns see GitHub.

| Comparison | # genes | Top 10 signaling pathways | -log(p-value) |
|---|---|---|---|
| LPS100 vs Control | 1779 | **Role of Macrophages, Fibroblasts and Endothelial Cells in Rheumatoid Arthritis** | 16.7 |
| | | Dendritic Cell Maturation | 12.9 |
| | | iNOS Signaling | 12.6 |
| | | **Role of IL-17A in Arthritis** | 12.5 |
| | | IL-10 Signaling | 11.8 |
| | | Role of Tissue Factor in Cancer | 11.6 |
| | | Production of Nitric Oxide and Reactive Oxygen Species in Macrophages | 11.1 |
| | | **Neuroinflammation Signaling Pathway** | 10.9 |
| | | Toll-like Receptor Signaling | 10.7 |
| | | CD40 Signaling | 10 |
| LPS100+A10 vs Control | 4721 | Molecular Mechanisms of Cancer | 14.1 |
| | | Colorectal Cancer Metastasis Signaling | 11.2 |
| | | Toll-like Receptor Signaling | 11 |
| | | **NF-κB Signaling** | 10.6 |
| | | **Role of Macrophages, Fibroblasts and Endothelial Cells in Rheumatoid Arthritis** | 10.4 |
| | | PI3K Signaling in B Lymphocytes | 10.4 |
| | | Role of Tissue Factor in Cancer | 10.3 |
| | | B Cell Receptor Signaling | 9.83 |

| | | TREM1 Signaling | 7.83 |
| | | Protein Kinase A Signaling | 7.81 |
| LPS100+B100 vs Control | 4049 | **Role of Macrophages, Fibroblasts and Endothelial Cells in Rheumatoid Arthritis** | 13.6 |
| | | **PI3K Signaling in B Lymphocytes** | 11.4 |
| | | Production of Nitric Oxide and Reactive Oxygen Species in Macrophages | 10.2 |
| | | Molecular Mechanisms of Cancer | 10.2 |
| | | Activation of IRF by Cytosolic Pattern Recognition Receptors | 10.2 |
| | | IL-10 Signaling | 9.63 |
| | | B Cell Receptor Signaling | 9.55 |
| | | iNOS Signaling | 9.21 |
| | | Role of Tissue Factor in Cancer | 9.06 |
| | | **Toll-like Receptor Signaling** | 8.93 |
| LPS100+A10 vs LPS100 | 8 | | |
| LPS100+B100 vs LPS100 | 0 | | |
| LPS100+A10 vs LPS100+B100 | 1132 | T Cell Exhaustion Signaling Pathway | 5.46 |
| | | Dendritic Cell Maturation | 5.43 |
| | | Role of Macrophages, Fibroblasts and Endothelial Cells in Rheumatoid Arthritis | 5.42 |
| | | Altered T Cell and B Cell Signaling in Rheumatoid Arthritis | 5.22 |
| | | Th17 Activation Pathway | 5.16 |
| | | Leukocyte Extravasation Signaling | 4.81 |
| | | Superpathway of Cholesterol Biosynthesis | 4.8 |
| | | IL-10 Signaling | 4.49 |
| | | Role of Tissue Factor in Cancer | 4.31 |

|  |  | Th1 and Th2 Activation Pathway | 4.26 |

Significant genes in microglia cultures at an FDR < 0.05 and a 1.5-fold change sorted on log(p-value).

**Table 5. Genes unique to LPS, agonistic and antagonistic stimulation of α7nAChR in single hit astrocytes cultures.**

|  | Top 10 signaling pathways | -log(p-value) |
|---|---|---|
| **Genes unique to LPS100** | Ephrin A Signaling | 3.21 |
|  | P2Y Purigenic Receptor Signaling Pathway | 2.72 |
|  | Role of p14/p19ARF in Tumor Suppression | 2.55 |
|  | Thiamin Salvage III | 2.2 |
|  | Melanoma Signaling | 2.14 |
|  | Lymphotoxin β Receptor Signaling | 2.03 |
|  | CCR3 Signaling in Eosinophils | 1.96 |
|  | CD40 Signaling | 1.84 |
| **Genes unique to A10** | Germ Cell-Sertoli Cell Junction Signaling | 9.13 |
|  | Integrin Signaling | 7.4 |
|  | Actin Cytoskeleton Signaling | 5.84 |
|  | Leukocyte Extravasation Signaling | 5.81 |
|  | Sertoli Cell-Sertoli Cell Junction Signaling | 5.58 |
|  | GDP-glucose Biosynthesis | 5.49 |
|  | Glucose and Glucose-1-phosphate Degradation | 5.24 |
|  | Phagosome Formation | 4.84 |
|  | Signaling by Rho Family GTPases | 4.71 |
|  | HGF Signaling | 4.6 |
| **Genes unique to B100** | Apelin Cardiomyocyte Signaling Pathway | 5.54 |

| | |
|---|---|
| Adrenomedullin signaling pathway | 4.09 |
| Dendritic Cell Maturation | 3.23 |
| Endothelin-1 Signaling | 3.2 |
| UVA-Induced MAPK Signaling | 3.2 |
| Renin-Angiotensin Signaling | 2.97 |
| Role of NFAT in Cardiac Hypertrophy | 2.95 |
| Phagosome Formation | 2.95 |
| GP6 Signaling Pathway | 2.91 |
| Wnt/Ca+ pathway | 2.89 |

# Experimental Design

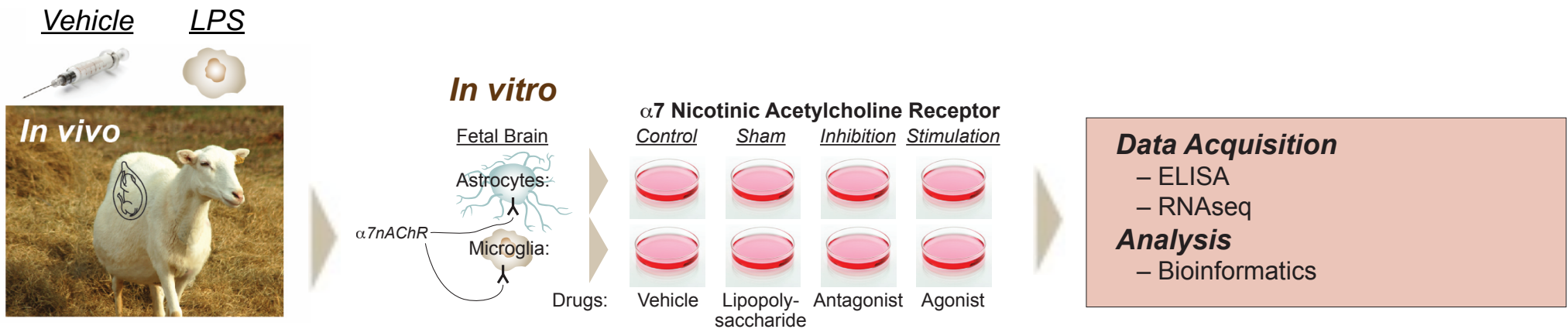

Cortes *et al.* J Neurosci Methods. 2017; 276: 23–32.

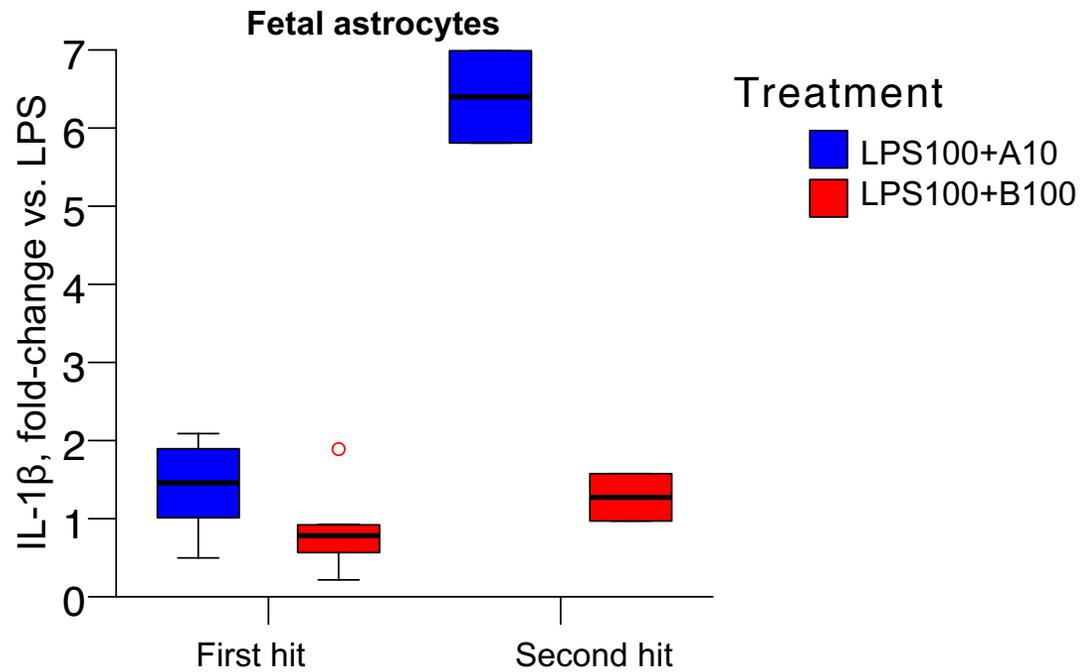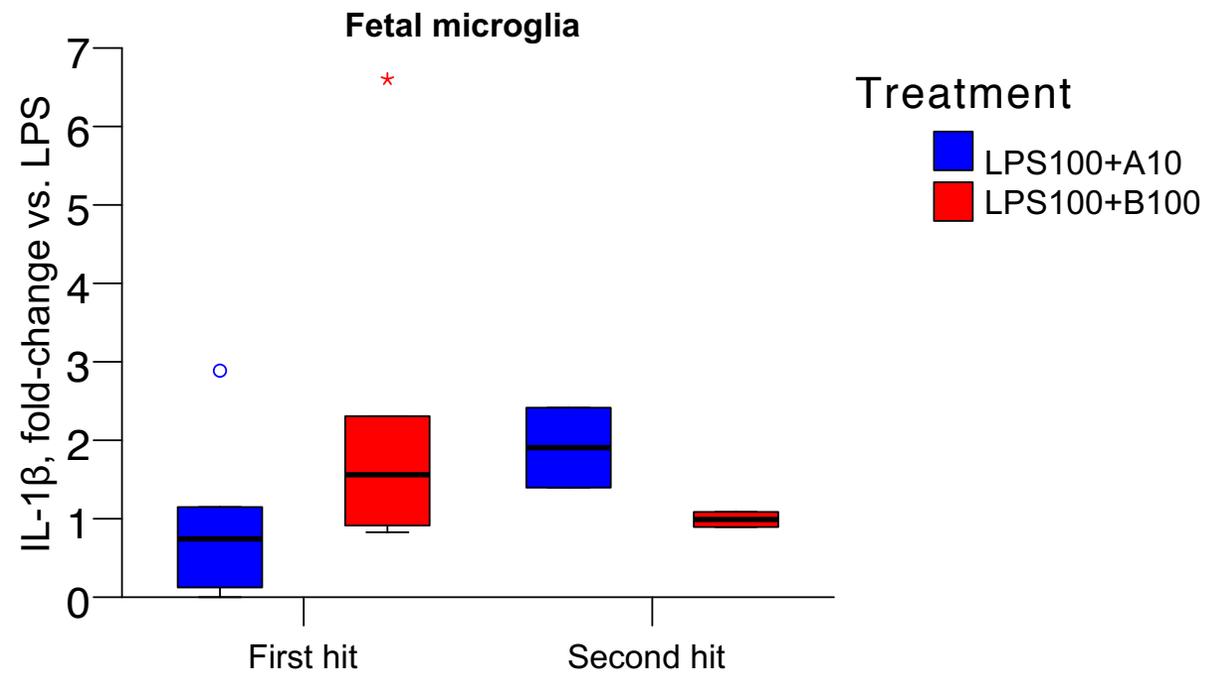

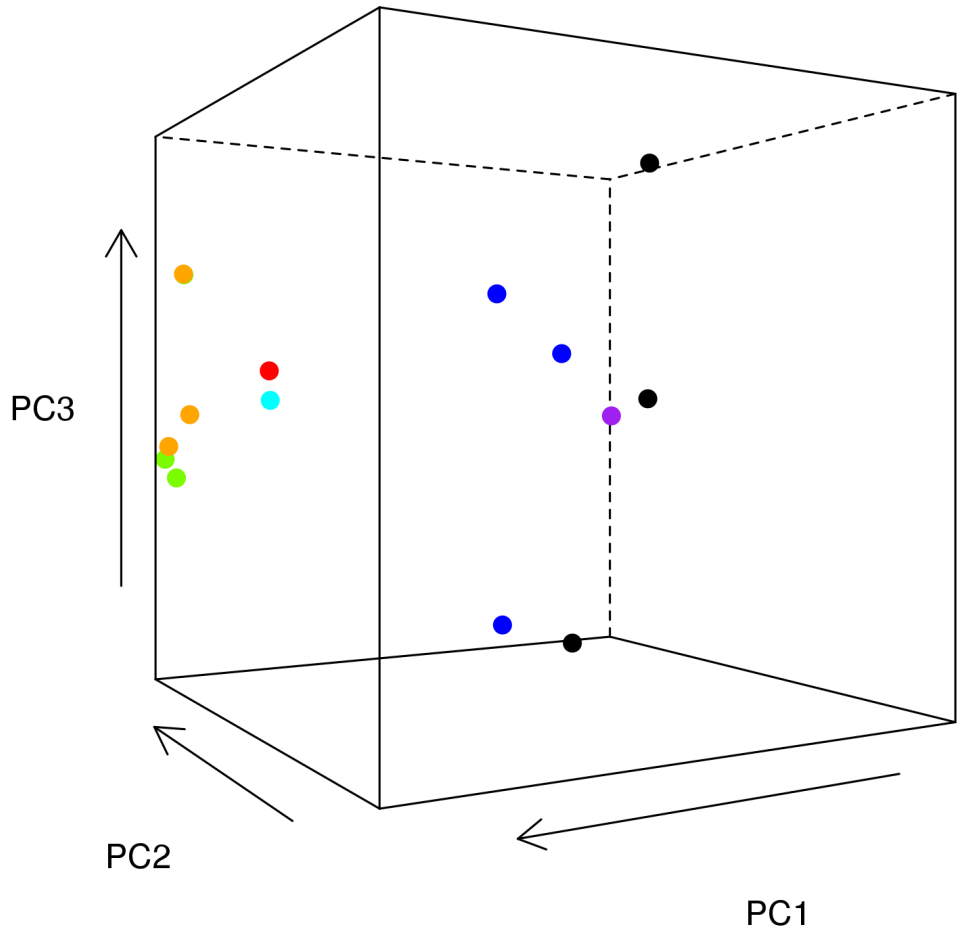

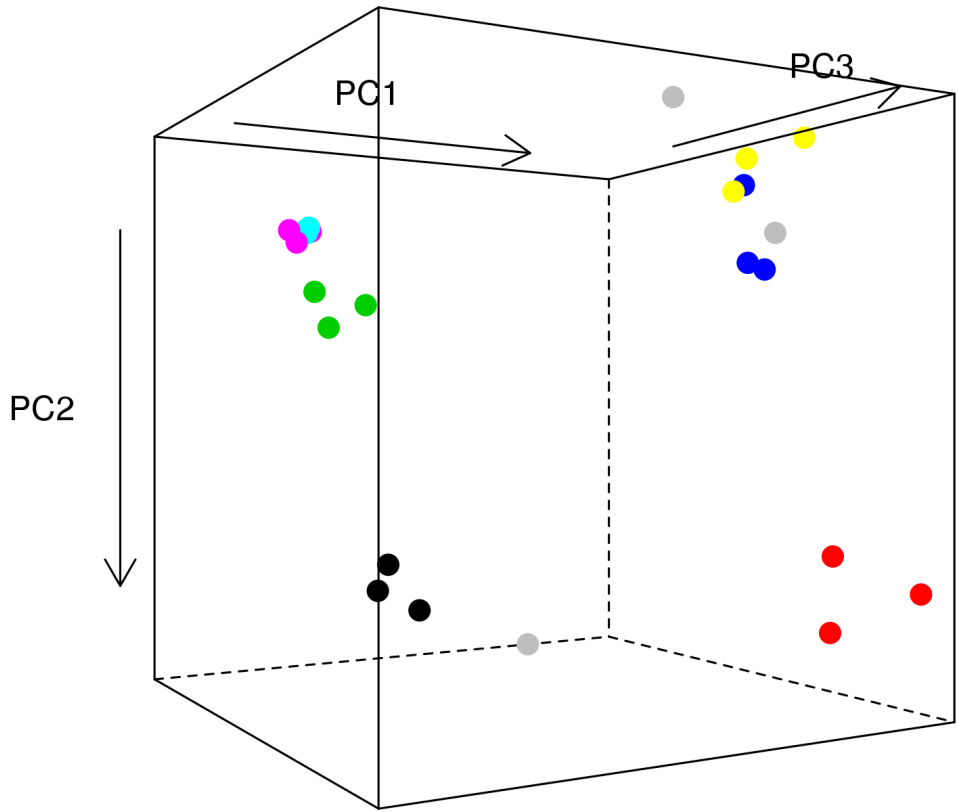

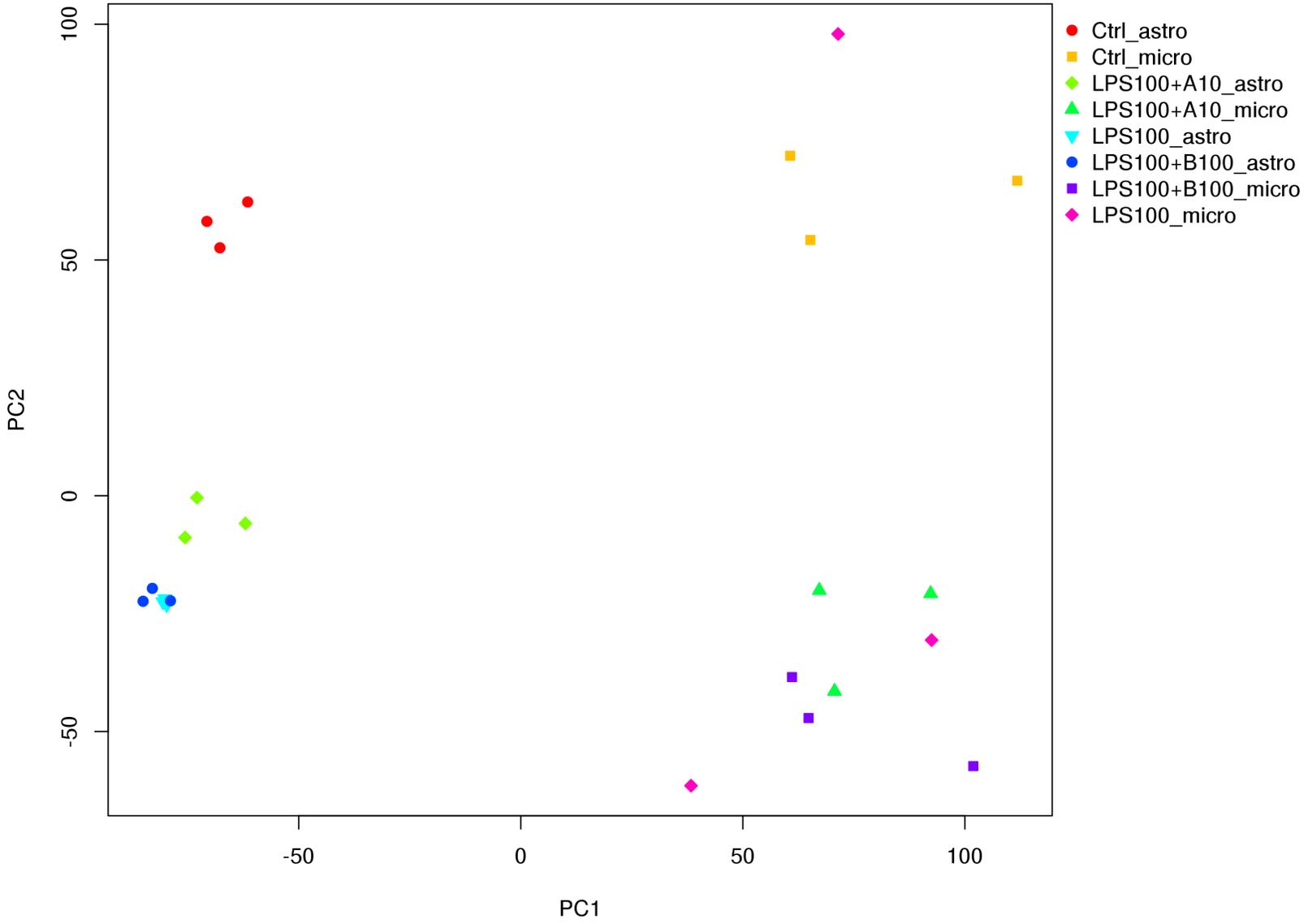